\documentclass[11pt]{article}
\usepackage{moriond,epsfig,amsmath,amssymb}
\newcommand{\abbrev}{\small}

\newcommand{\eqn}[1]{Eq.\,(\ref{#1})}
\newcommand{\fig}[1]{Fig.\,\ref{#1}}

\newcommand{\dd}{{\rm d}}

\newcommand{\lo}{{\abbrev LO}}
\newcommand{\nlo}{{\abbrev NLO}}
\newcommand{\nnlo}{{\abbrev NNLO}}
\newcommand{\muF}{\mu_{F}}
\newcommand{\muR}{\mu_{R}}
\newcommand{\sm}{{\abbrev MSSM}}
\newcommand{\mssm}{{\abbrev MSSM}}
\newcommand{\mhiggs}{M_H}
\newcommand{\lhc}{{\abbrev LHC}}
\begin{document}
\vspace*{2cm}
\begin{flushright}
hep-ph/0305204 --- {\small BNL-HET}-03/11 --- {\small CERN-TH}/2003-112
--- May 2003
\end{flushright}
\vspace*{2cm} \title{Techniques for NNLO Higgs production in the
  Standard Model and the MSSM
  }

\author{\underline{R.V. HARLANDER}${}^1$ and W.B. KILGORE${}^2$}

\address{
  ${}^{1)}$TH Division, CERN, CH-1211 Geneva 23, Switzerland\\
  ${}^{2)}$Department of Physics, Brookhaven National Lab,
  Upton, New York 11973, USA
  }

\maketitle\abstracts{ New techniques developed in connection with the
  {\footnotesize NNLO} corrections to the Higgs production rate at
  hadron colliders and some recent applications are reviewed.  }
%- }}}
%- {{{ Introduction:

\section{Introduction}
%----------------------------------------------------------------------

The \nlo{} corrections for the dominant Higgs production mechanism at
hadron colliders, $gg\to H$, amount to about 70$\%$ and suffer from
rather large scale
uncertainties.\cite{Dawson:1990zj,Djouadi:1991tk,Graudenz:1992pv} The
need for the evaluation of the \nnlo{} cross section has resulted in
promising new calculational techniques. The first part of this talk will
briefly review these techniques. In the second part, we will discuss a
recent application, namely the \nnlo{} cross section for \mssm{} Higgs
production in bottom quark fusion.

\section{Techniques for Higgs production at NNLO}
The first calculation of the \nnlo{} prediction for the cross section
$\sigma(pp\to H+X)$ used the classic approach of computing the
amplitudes for virtual and real corrections, squaring them, and
integrating over the final state phase space.  The two-loop virtual
amplitude for $gg\to H$ was evaluated\,\cite{Harlander:2000mg} using the
method of Baikov and Smirnov\,\cite{Baikov:2000jg} that maps the
occuring integrals to the well-known class of three-loop propagator-type
integrals.\cite{Chetyrkin:1981qh} The phase space integration for the
two-loop virtual terms is trivial, resulting in $\delta(1-x)$, where
$x=\mhiggs^2/\hat s$ and $\hat s$ is the partonic center of mass (c.m.)\ 
energy.

The one-loop amplitude for the radiation of a single massless parton has
to be interfered with the corresponding tree-level expression, and
integrated over the two-particle phase space. Both loop and phase space
integration can be performed analytically.

This leaves us with the tree-level contributions for the radiation of
two massless partons. The squared amplitude can be obtained
straightforwardly with the help of {\tt FORM}.\cite{Vermaseren:2000nd}
In the first approach that we are going to describe, the phase space
integrals were evaluated in terms of an expansion in $(1-x)$. The
leading term is called the {\it soft
  approximation}\,\cite{Harlander:2001is,Catani:2001ic} and is formally
of order $(1-x)^{-1}$, where the associated divergence as $x\to 1$ is
parameterized in terms of the distributions $\delta(1-x)$ and
$[\ln^n(1-x)/(1-x)]_+$.  The higher orders in $(1-x)$ can be obtained by
a systematic expansion of the squared amplitude and the phase space
measure.\cite{Harlander:2002wh} The crucial point is that, independent
of the degree of this expansion, one always ends up with the same type
of integrals. This classifies the procedure as an algorithm, which can
be fully automated.

Integrating the resulting expansion of the partonic cross section over
the parton densities, one observes that higher orders in $(1-x)$ are
numerically irrelevant, and the resulting hadronic cross section is
phenomenologically equivalent to the result derived from the exact
partonic cross section.\cite{Harlander:2002wh} On the other hand, one
can make an ansatz of a sum of polylogarithms with unknown coefficients,
expand it in $(1-x)$, and compare the result with the expansion obtained
for the partonic cross section.\cite{Kilgore:2002sk} Given that this
expansion is known to sufficiently high (but finite!)  orders, this
determines the coefficients of the polylogarithms and thus the exact
result for the partonic cross
section.\cite{Kilgore:2002sk,Harlander:2002vv}

Clearly, the method of phase-space expansion is a bottom-up approach:
starting at the soft approximation, one can successively improve the
accuracy of the result by including higher orders in $(1-x)$, until a
sufficient number of terms is known to invert the series and arrive at
the exact result.

A second method to obtain the \nnlo{} result for the Higgs production
rate has been developed by Anastasiou and
Melnikov\,\cite{Anastasiou:2002yz}. According to the Cutkosky rules, one
can write, for example,
\vspace{-.5em}
\begin{equation}
\begin{split}
%  \int\dd{\rm PS}\,\bigg|
%  \epsfxsize=8em
%  \raisebox{-2.7em}{\epsffile[160 450 465 665]{figs/nnlo_2real1_1.ps}}
%  \bigg|^2 \ =\
%  \epsfxsize=8.5em
%  \raisebox{-3.5em}{\epsffile[110 490 400 740]{figs/nnlo_cutkoski1_1.ps}}
%  \int\dd{\rm PS}\,\bigg|
%  \epsfxsize=6em
%  \raisebox{-2.2em}{\epsffile[160 450 465 665]{figs/nnlo_2real1_1.ps}}
%  \bigg|^2 \ =\
%  \epsfxsize=6.5em
%  \raisebox{-2.6em}{\epsffile[110 490 400
%  740]{figs/nnlo_cutkoski1_1.ps}}\\[-2.5em]
  \int\dd{\rm PS}\,\bigg|
  \epsfxsize=6.5em
  \raisebox{-2.2em}{\epsffile[160 450 465 665]{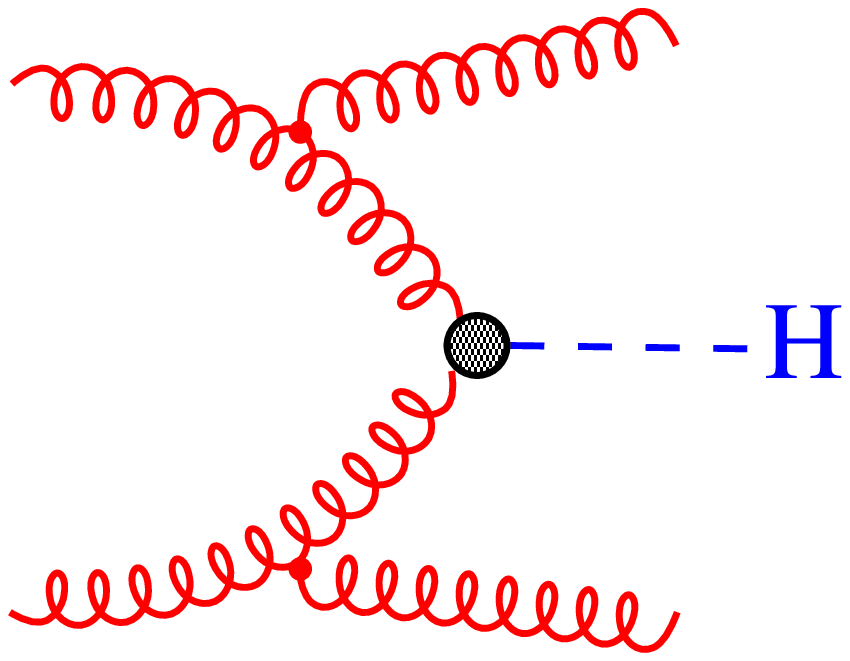}}
  \bigg|^2 \ =\
  \epsfxsize=7em
  \raisebox{-2.8em}{\epsffile[110 490 400
  740]{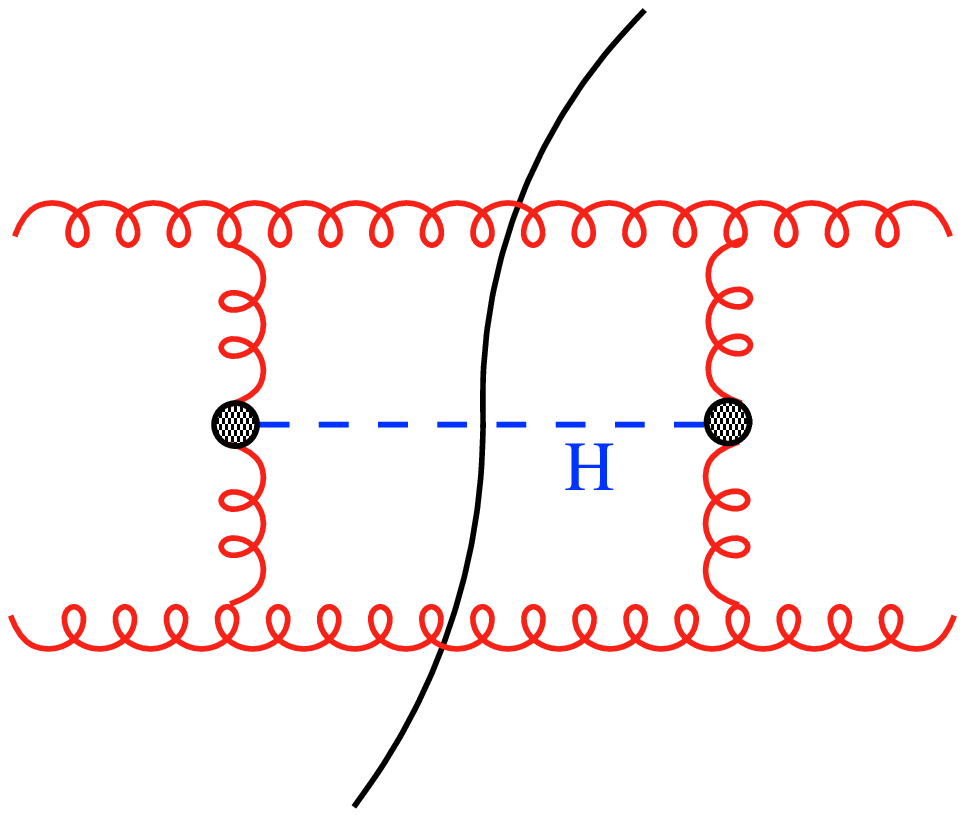}}\\[-2.5em]
  \label{eq::cut}
\end{split}
\end{equation}
where the initial and final states of the diagram on the r.h.s.\ are
identical. {\it Without cuts}, this diagram would be a double-box
diagram with two external scales, $\hat s$ and $M_H$. Such diagrams can
be evaluated with the help of the general algorithms that have been
developed within the last few years in the context of $2\to 2$
scattering amplitudes at \nnlo{} (see, e.g., a recent
review\,\cite{Glover:2002gz}). The crucial observation of Anastasiou and
Melnikov\,\cite{Anastasiou:2002yz} was that these algorithms are
directly applicable to {\it cut diagrams} of the kind shown on the
r.h.s.\ of \eqn{eq::cut}.\footnote{Let us remark that the meaning of
  ``cut'' in this context is not restricted to Cutkosky-cuts. For
  example, the cut lines can be initial rather than final states (which
  leads to the method used to originally derive the virtual
  terms\,\cite{Baikov:2000jg,Harlander:2000mg}), or they can be
  restricted to a particular kinematic
  configuration.\cite{Anastasiou:2002qz}} In this way, the partonic
cross section for Higgs production was derived in closed
form\,\cite{Anastasiou:2002yz}. Needless to say that a re-expansion of
this closed expression recovered the expressions derived through
phase-space expansion\,\cite{Harlander:2002wh}. In turn, the result
obtained by inverting the
expansion\,\cite{Kilgore:2002sk,Harlander:2002vv} confirmed the closed
expression. The \nnlo{} results for the production of a pseudo-scalar
Higgs were obtained independently and simultaneously in both
approaches\,\cite{Harlander:2002vv,Anastasiou:2002wq}.
Meanwhile, the \nnlo{} results for both scalar and pseudo-scalar Higgs
production have been re-confirmed using the analoguous approach that was
applied to the \nnlo{} Drell-Yan calculation.\cite{Ravindran:2003um} 

The phenomenological implications of the \nnlo{} are significant and
have been discussed extensively in the
literature.\cite{Harlander:2002wh,Anastasiou:2002yz,Ravindran:2003um}
They shall not be repeated here due to space limitations.

%- }}}
%- {{{ Bottom fusion:

\section{Higgs production in bottom quark fusion}
The production rate for a {\abbrev SM} Higgs boson being under good
theoretical control, one may ask to which extent these results are
applicable also for Higgs boson production in supersymmetric models. For
simplicity, we will restrict ourselves to the Minimal Supersymmetric
Standard Model (\mssm{}) in this talk.

The answer is that the \nnlo{} production rate of a neutral scalar Higgs
boson in the \mssm{} can be inferred directly from the known \sm{}
prediction in only a rather restricted parameter space, in particular
for small $\tan\beta$ and large squark masses. Also the production of a
pseudo-scalar Higgs boson has been evaluated at
\nnlo{}\,\cite{Harlander:2002vv,Anastasiou:2002wq,Ravindran:2003um}
within these restrictions.

In other regions of parameter space, virtual contributions of
sypersymmetric particles such as squarks may become
important.\cite{Dawson:1996xz} On the other hand, large values of
$\tan\beta$ enhance the Yukawa coupling of bottom quarks. Thus, the
$gg\phi$ coupling may have a significant contribution from virtual
bottom quarks ($\phi$ denotes any of the neutral Higgs bosons in the
\mssm{}). In this case, the \nnlo{} corrections are much harder to
evaluate, because the effective-Lagrangian approach of the top quark
case is not expected to work.

The main focus here shall be another effect of an enhanced bottom Yukawa
coupling, namely the increased rate of associated production of a Higgs
boson with a bottom--anti-bottom quark pair. There has been an on-going
discussion concerning the proper description of this
process.\cite{Spira:2002rd,Plehn:2002vy,Boos:2003yi,Maltoni:2003pn} {\it
  A priori}, the leading order contribution is, of course, the
tree-level process $gg\to \phi b\bar b$, where $\phi$ denotes any of the
\mssm{} Higgs bosons $h,H,A$.  However, when evaluating the total rate
for this process, the integration over small bottom-$p_T$ leads to
collinear logarithms of the form $l_b \equiv \ln(m_b^2/M^2)$, where $M$
is a scale of the order of the Higgs boson mass. Since $m_b\ll
M_\phi\sim M$, these logarithms should be resummed.  This is achieved by
introducing bottom quark densities and making the process $b\bar b\to
\phi$ the leading order contribution.  Schematically, one can write the
total cross section in the bottom density approach as follows:
\begin{equation}
\begin{split}
\sigma(pp\to H+X) = \sum_{n=0}^\infty 
(\alpha_s l_b)^{n+2}
\left[ c_{n0} + c_{n1}\frac{1}{l_b} 
  + \frac{1}{l_b^2}\left( c_{n2} + \alpha_s d_{n3} + \alpha_s^2\,d_{n4}
  + \ldots\right)\right]\,.
\label{eq::schematic}
\end{split}
\end{equation}
This equation is to be understood as follows: First of all, one should
note that the $c_{ni}$ and $d_{ni}$ are not obtained individually for
each $n$, because the sum over $n$ is implicit in the parton densities.
Including only the leading order process $b\bar b\to \phi$, one obtains
the contribution from the $c_{n0}$.  The \nlo{} diagrams contribute
terms of order $1/l_b$ (e.g., $gb\to b\phi$) and of order $\alpha_s$
($b\bar b\to \phi$ at 1-loop) with respect to the leading term. Both are
contained in the $c_{n1}$.  At {\abbrev N$^n$LO}, for $n\geq 2$, we have
terms of order $\alpha_s^{n-k}/l_b^k$ w.r.t.\ \lo{}, where $k=0,1,2$.
The reason why there can be only two inverse powers of $l_b$ comes from
the fact that there are only two initial state partons.  Looking at
\eqn{eq::schematic}, it becomes clear why the \nnlo{} plays an
exceptional role in this process: It comprises all terms at leading
order in $\alpha_s$.

The calculation of the process\,\footnote{We adopt the notation $(b\bar
  b)\phi$ in order to indicate that the bottom quarks may be produced at
  small transverse momenta and thus escape detection.} $pp\to (b\bar
b)\phi$ in the bottom density approach proceeds in complete analogy to,
say, Drell-Yan production of virtual photons. Technical details of the
calculation can be found elsewhere.\cite{Harlander:2003ai}

\begin{figure}
  \begin{center}
  \begin{tabular}{cc}
    \mbox{\hspace{14em}}$(a)$ & \mbox{\hspace{17em}}$(b)$ \\[-1em]
    \epsfxsize=5.5cm
    \epsffile[110 265 465 560]{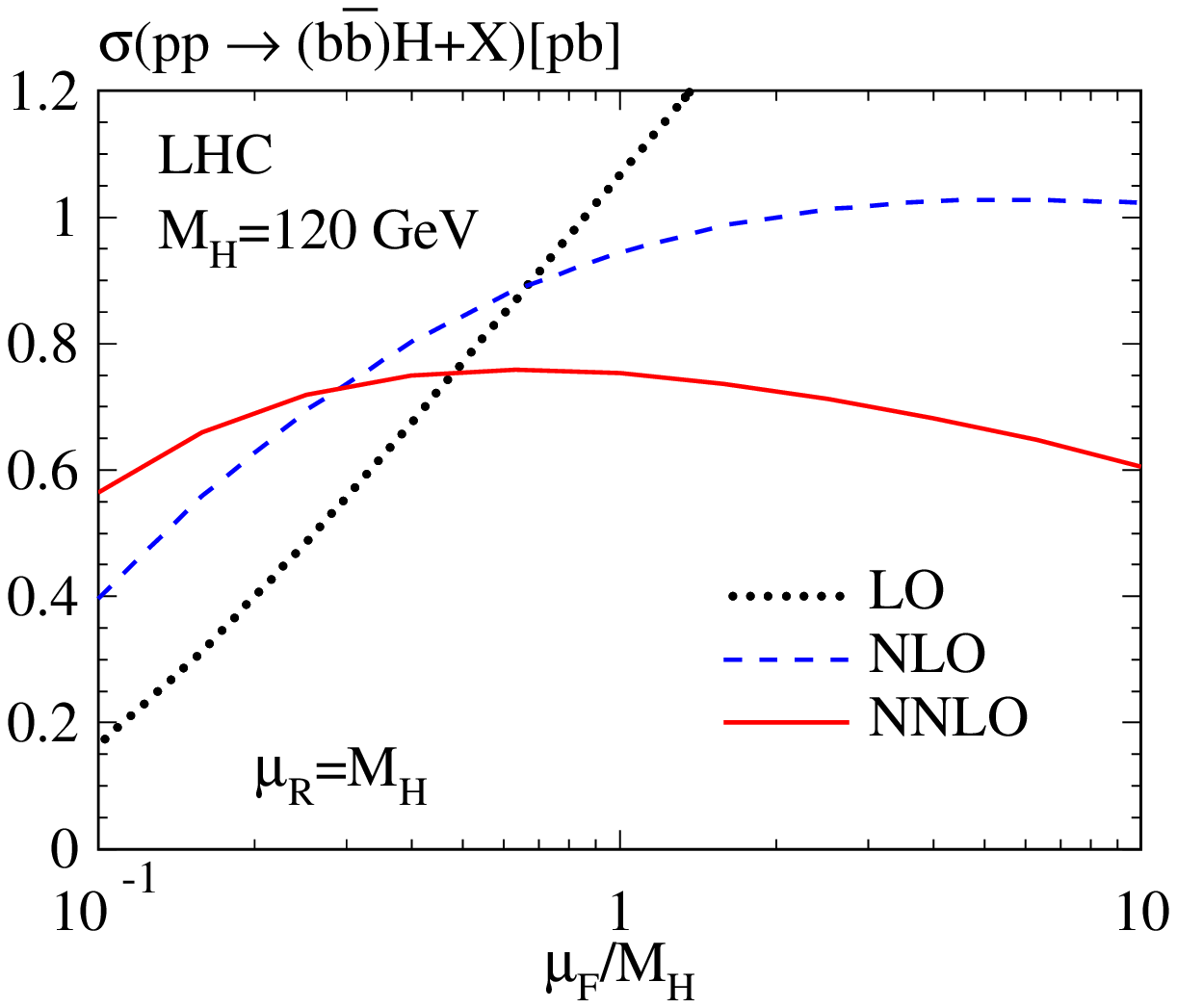} &\hskip 1cm
    \epsfxsize=5.5cm
    \epsffile[110 265 465 560]{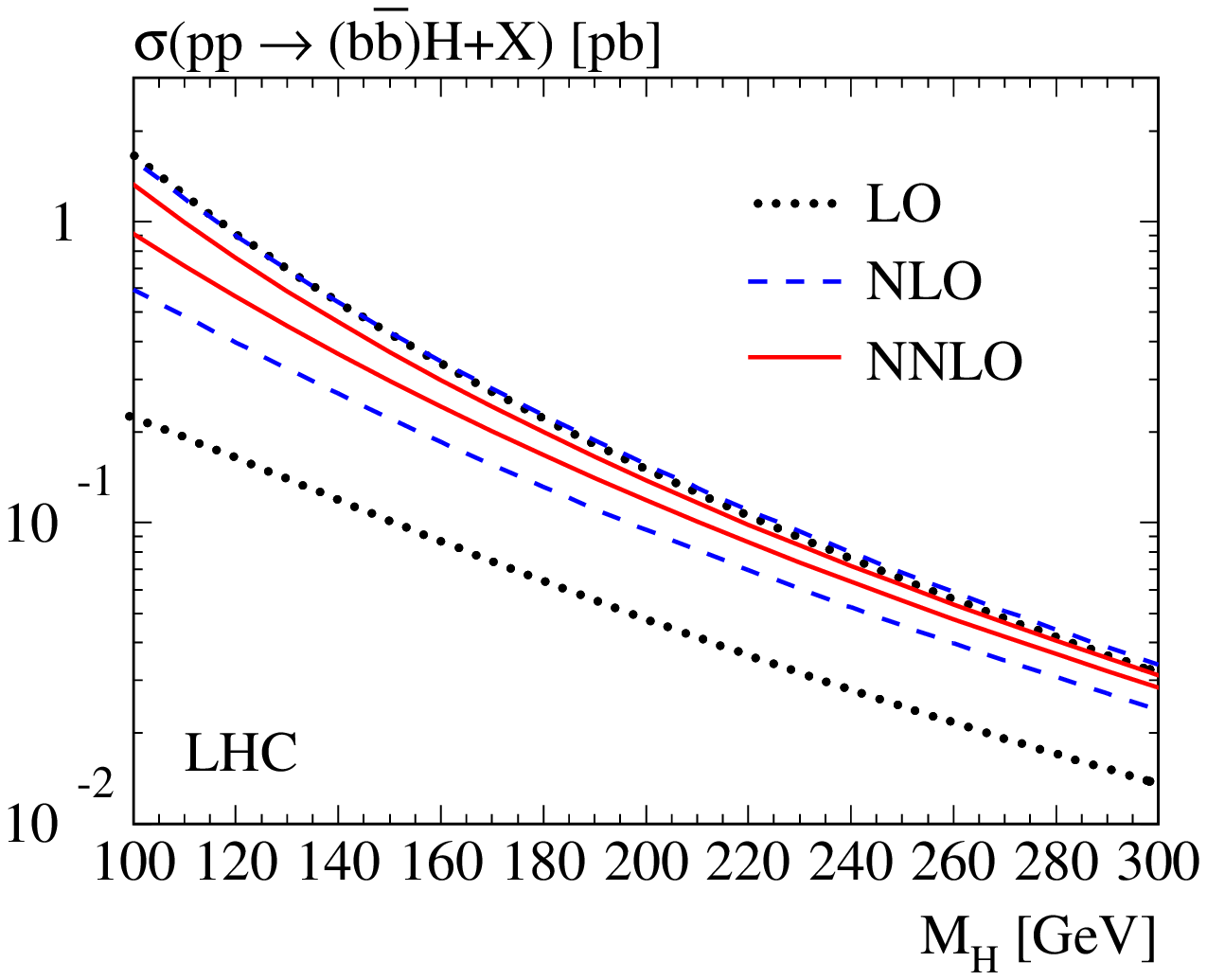}
  \end{tabular}
\end{center}
\caption[]{\label{fig::xsec}} $(a)$ Factorization scale dependence of
  the cross section for $pp\to (b\bar b)H+X$ ($\tan\beta=1$); $(b)$
  Total cross section for $pp\to (b\bar b)H+X$ --- upper/lower line:
  $\muF=0.7M_H$/$\muF=0.1M_H$ [$\muR=M_H$].
\end{figure}

\fig{fig::xsec}\,$(a)$ shows the factorization scale dependence of the
cross section at \lo{}, \nlo{}, and \nnlo{} at the \lhc{}, for a Higgs
mass of $M_H=120$\,GeV at the {\abbrev LHC}. We notice several
intriguing features: First, in contrast to the \lo{} and \nlo{} result,
the \nnlo{} curve has a clearly distinguished point where the derivative
is zero, {\it i.e.} where the sensitivity to $\muF$ is minimal. Second,
the \nnlo{} corrections are zero at a point where the \nlo{} corrections
are small. Third, this point is close to the point of least sensitivity.
And fourth, all these points are compatible with a previous
estimate\,\cite{Plehn:2002vy,Maltoni:2003pn,Boos:2003yi} of the
``natural'' factorization scale for this process of around $\muF=M_H/4$.
%Another striking fact\,\cite{Harlander:2003ai} (not shown here due to
%space limitations) is that for $\muF=M_H/4$, the renormalization scale
%dependence of the total cross section is very small, in contrast to the
%choice of, say, $\muF=M_H$. 
For the Tevatron, the overall picture is
essentially the same\,\cite{Harlander:2003ai}.  These features nicely
demonstrate the self-consistency of the bottom density approach and
support the general considerations concerning the proper choice of the
factorization scale for heavy quark
partons.\cite{Plehn:2002vy,Maltoni:2003pn,Boos:2003yi}

\fig{fig::xsec}\,$(b)$ shows the total cross section for $pp\to (b\bar
b)H+X$ as a function of the Higgs boson mass up to \nnlo{}, for two
different values of the factorization scale, indicating the theoretical
uncertainty (the renormalization scale dependence can be neglected).

\paragraph{Conclusions.} Recent progress in the evaluation of radiative
corrections has led to \nnlo{} predictions for Higgs production at
hadron colliders, both in the {\abbrev SM} and the {\abbrev MSSM}. The
results are very stable with respect to scale variations and indicate a
well-behaved perturbative series.

%- }}}
%- {{{ bibliography:

\def\app#1#2#3{{\it Act.~Phys.~Pol.~}\jref{\bf B #1}{#2}{#3}}
\def\apa#1#2#3{{\it Act.~Phys.~Austr.~}\jref{\bf#1}{#2}{#3}}
\def\annphys#1#2#3{{\it Ann.~Phys.~}\jref{\bf #1}{#2}{#3}}
\def\cmp#1#2#3{{\it Comm.~Math.~Phys.~}\jref{\bf #1}{#2}{#3}}
\def\cpc#1#2#3{{\it Comp.~Phys.~Commun.~}\jref{\bf #1}{#2}{#3}}
\def\epjc#1#2#3{{\it Eur.\ Phys.\ J.\ }\jref{\bf C #1}{#2}{#3}}
\def\fortp#1#2#3{{\it Fortschr.~Phys.~}\jref{\bf#1}{#2}{#3}}
\def\ijmpc#1#2#3{{\it Int.~J.~Mod.~Phys.~}\jref{\bf C #1}{#2}{#3}}
\def\ijmpa#1#2#3{{\it Int.~J.~Mod.~Phys.~}\jref{\bf A #1}{#2}{#3}}
\def\jcp#1#2#3{{\it J.~Comp.~Phys.~}\jref{\bf #1}{#2}{#3}}
\def\jetp#1#2#3{{\it JETP~Lett.~}\jref{\bf #1}{#2}{#3}}
\def\jhep#1#2#3{{\small\it JHEP~}\jref{\bf #1}{#2}{#3}}
\def\mpl#1#2#3{{\it Mod.~Phys.~Lett.~}\jref{\bf A #1}{#2}{#3}}
\def\nima#1#2#3{{\it Nucl.~Inst.~Meth.~}\jref{\bf A #1}{#2}{#3}}
\def\npb#1#2#3{{\it Nucl.~Phys.~}\jref{\bf B #1}{#2}{#3}}
\def\nca#1#2#3{{\it Nuovo~Cim.~}\jref{\bf #1A}{#2}{#3}}
\def\plb#1#2#3{{\it Phys.~Lett.~}\jref{\bf B #1}{#2}{#3}}
\def\prc#1#2#3{{\it Phys.~Reports }\jref{\bf #1}{#2}{#3}}
\def\prd#1#2#3{{\it Phys.~Rev.~}\jref{\bf D #1}{#2}{#3}}
\def\pR#1#2#3{{\it Phys.~Rev.~}\jref{\bf #1}{#2}{#3}}
\def\prl#1#2#3{{\it Phys.~Rev.~Lett.~}\jref{\bf #1}{#2}{#3}}
\def\pr#1#2#3{{\it Phys.~Reports }\jref{\bf #1}{#2}{#3}}
\def\ptp#1#2#3{{\it Prog.~Theor.~Phys.~}\jref{\bf #1}{#2}{#3}}
\def\ppnp#1#2#3{{\it Prog.~Part.~Nucl.~Phys.~}\jref{\bf #1}{#2}{#3}}
\def\rmp#1#2#3{{\it Rev.~Mod.~Phys.~}\jref{\bf #1}{#2}{#3}}
\def\sovnp#1#2#3{{\it Sov.~J.~Nucl.~Phys.~}\jref{\bf #1}{#2}{#3}}
\def\sovus#1#2#3{{\it Sov.~Phys.~Usp.~}\jref{\bf #1}{#2}{#3}}
\def\tmf#1#2#3{{\it Teor.~Mat.~Fiz.~}\jref{\bf #1}{#2}{#3}}
\def\tmp#1#2#3{{\it Theor.~Math.~Phys.~}\jref{\bf #1}{#2}{#3}}
\def\yadfiz#1#2#3{{\it Yad.~Fiz.~}\jref{\bf #1}{#2}{#3}}
\def\zpc#1#2#3{{\it Z.~Phys.~}\jref{\bf C #1}{#2}{#3}}
\def\ibid#1#2#3{{ibid.~}\jref{\bf #1}{#2}{#3}}
\newcommand{\bibentry}[4]{#1, #3.}
\newcommand{\arxiv}[1]{{\tt #1}}
\newcommand{\jref}[3]{{\bf #1}, #3 (#2)}

%- }}}

\end{document}